# Bioinspired Nanocomposites: 2D Materials Within a 3D Lattice

*Matteo Di Giosia, Iryna Polishchuk, Eva Weber, Simona Fermani, Luca Pasquini, Nicola M. Pugno, Francesco Zerbetto, Marco Montalti, Matteo Calvaresi, Giuseppe Falini,\* Boaz Pokroy\**

M. Di Giosia,[+] Dr. S. Fermani, Prof. F. Zerbetto, Prof. M. Montalti, Prof. M. Calvaresi, Prof. G. Falini
Dipartimento di Chimica "Giacomo Ciamician", Alma Mater Studiorum,
Università di Bologna. Via F. Selmi, 2
40126 Bologna, Italy
E-mail: giuseppe.falini@unibo.it

Dr. I. Polishchuk,[+] Dr. E. Weber, Prof. B. Pokroy
Department of Materials Science and Engineering and the Russell Berrie Nanotechnology Institute, Technion—Israel Institute of Technology
32000 Haifa, Israel
E-mail: bpokroy@tx.technion.ac.il

Prof. L. Pasquini
Dipartimento di Fisica e Astronomia, Alma Mater Studiorum
Università di Bologna, Viale Berti Pichat 6/2,
40126 Bologna, Italy

Prof. N. Pugno
Laboratory of Bio-inspired & Graphene Nanomechanics, Department of Civil, Environmental and Mechanical Engineering, University of Trento, via Mesiano 77,
38123 Trento, Italy
Center for Materials and Microsystems, Fondazione Bruno Kessler, via Sommarive 18,
38123 Povo, Italy
School of Engineering & Materials Science, Queen Mary University of London, Mile End Road, London E1 4NS, United Kingdom

## Abstract

Composites, materials composed of two or more materials—metallic, organic, or inorganic—usually exhibit the combined physical properties of their component materials. The result is a material that is superior to conventional monolithic materials. Advanced composites are used in a variety of industrial applications and therefore attract much scientific interest. Here we describe the formation of novel carbon-based nanocomposites via incorporation of graphene oxide (GO) into the crystal lattice of single crystals of calcite. Incorporation of a 2D organic material into single-crystal lattices has never before been reported. To characterize the resulting nanocomposites we employed high-resolution synchrotron powder X-ray diffraction, electron microscopy, transmission electron microscopy, fluorescence microscopy and nanoindentation tests. Detailed analysis revealed a layered distribution of GO sheets incorporated within the calcite host. Moreover, the optical and mechanical properties of the calcite host were altered when a carbon-based nanomaterial was introduced into its lattice. Compared to pure calcite, the composite GO/calcite crystals exhibited lower elastic modulus and higher hardness. The results of this study show that incorporation of a 2D material within a 3D crystal lattice is not only feasible but also can lead to the formation of hybrid crystals exhibiting new properties.

# 1. Introduction

Many materials formed in the process of biomineralization are nanocomposites. [1-3] Prominent among these are $CaCO_3$-based biominerals, found in organisms such as sea urchin skeleton nacre in mollusk shells, brittle star arm plates and shrimps. In these hard tissues the inclusion of biological macromolecules within the growing $CaCO_3$ mineral phase can generate complex composites with enhanced properties.[4-9] Intracrystalline organic macromolecules incorporated within a calcite single-crystal host have been found responsible for the sophisticated morphologies and the increased fracture toughness of these biogenic calcite crystals.[10-14] The incorporated organic matter induces the pronounced lattice distortions and distinctive microstructures observed in them.[15-17]

The biological concept underlying crystal formation has been successfully adapted in different synthetic systems to fabricate bioinspired materials with extraordinary characteristics. Interestingly, not only macromolecules[18] but also nanoparticles can become incorporated into calcite single crystals.[10, 19-24] Carbon nanoparticles, in particular, strongly affect the $CaCO_3$ crystallization process, thereby facilitating the formation of advanced composite materials.[22, 23] 2D materials such as graphene can provide a basis for further investigation of the incorporation of carbon-based molecules into the calcite host. Because of its distinctive mechanical features (high strength, toughness) and physicochemical properties (high surface area, electrical and thermal conductivities, low weight), graphene is a promising material for the fabrication of novel carbon-based nanocomposites.[25-27]

The mechanical properties of graphene/inorganic composites have been found to be significantly superior to those of their inorganic monolithic counterparts.[28-33] Nevertheless, the incorporation of such 2D materials into single-crystal lattices has never been reported. Here we report the incorporation of a 2D material, graphene oxide (GO), into the crystalline lattice of singe crystals of calcite. We selected GO as a 2D material model because the exposed surfaces of both sides of GO sheets exhibit functional groups capable of interacting with $Ca^{2+}$. This should potentially allow for effective interaction between GO sheets and calcite during the crystallization process. It has been previously shown that reduced GO (rGO) sheets interact with the $CaCO_3$.[34] Moreover, the relatively high surface area of this 2D material contributes to its high surface energy. This latter feature can act as a driving force for the incorporation of GO into a 3D crystal owing to significant reduction of the interfacial Gibbs free energy during the crystallization process.[35]

In addition, GO has interesting optical[36] and mechanical[37] properties not possessed by the calcite host. GO is known for its high elastic modulus and strength and also for its fluorescence properties. With regard to its optical characteristics, GO exhibits excitation energy-dependent fluorescence in water with emission peak maxima ranging from the red to the near-infrared spectral range.[38-44] Time-resolved fluorescence measurements of graphene oxide in water show multiexponential decay kinetics with lifetimes from 1 ps to 2 ns.[38]

## 2. Results and Discussion

The hybrid GO/calcite crystals that we grew were characterized structurally, spectroscopically and mechanically in order to investigate the mechanism by which the intercalation of a 2D material into a 3D lattice alters the chemical and physical properties of both the host and the guest.

First, we developed an optimal procedure for entrapping GO sheets into the single crystals of calcite (see Experimental section). Once the hybrid crystals were successfully grown we used high-resolution scanning electron microscopy (HRSEM) to study their morphologies. Figure 1 shows some GO sheets and flakes emerging from the hybrid crystals (see arrows in Figure 1B, D). This was confirmed by energy-dispersive X-ray spectroscopy (EDS) performed directly on those regions that are clearly carbon rich (35.55±0.62wt% of carbon as compared to 8.74±0.42wt% on the regions without GO flakes). In contrast to the sharp edges of the pure calcite crystals that serve as a control (Figure 1A), the GO/calcite crystals demonstrate a layered structure, possibly indicating a layered distribution of the GO sheets entrapped within the crystal (Figure 1B, D).

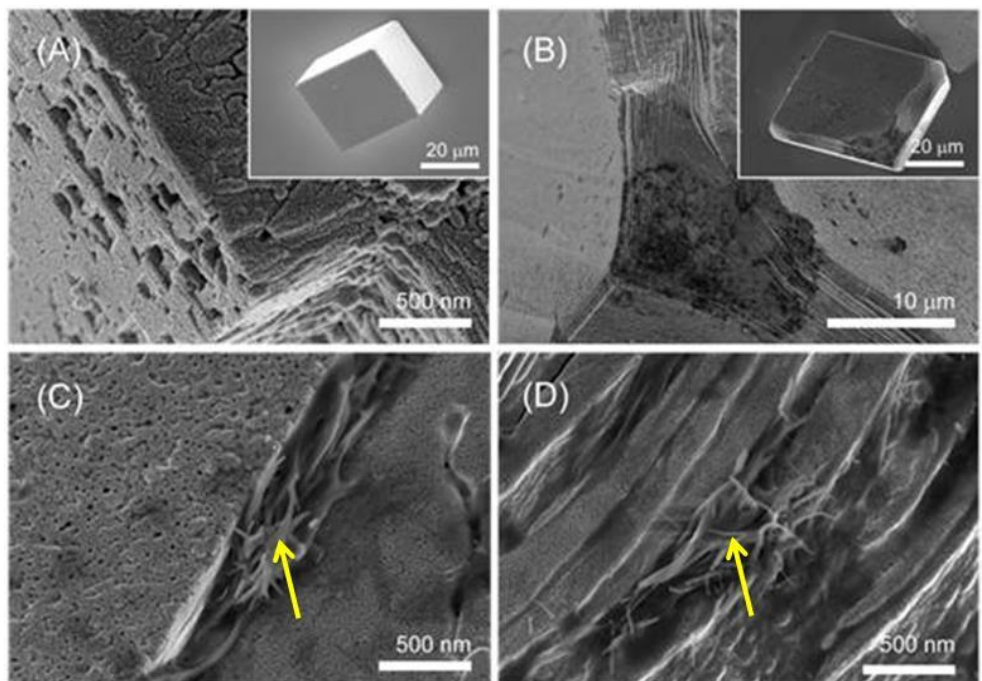

**Figure 1.** Representative HRSEM images of calcite single crystals. (A) Pure calcite crystal grown in the absence of an additive. The crystal exhibits a stepped surface as a result of the extensive washing process in deionized water that gently etches the surface. (B−D) GO/Calcite hybrid single crystal imaged at different magnifications on the surface (C) and at the edge regions (B−D). Insets show corresponding images at low magnification. Arrows indicate GO flakes emerging from the crystal.

The crystal structure of GO/calcite crystals at the atomic scale was further studied by means of state-of-the-art high-resolution aberration-corrected transmission electron microscopy (HRTEM). Our findings indicated that individual calcite crystals grown in the presence of GO are indeed single crystalline. HRTEM image (Figure 2A) demonstrates lattice fringes with a $d$-spacing of 3.9Å which corresponds to the {012} planes. The inset in Figure 2A displays the fast Fourier transforms (FFT) of the coresponding TEM image. As can be seen the FFT generated fits that of a single crystal (inset in Figure 2A) . To investigate the apparent distribution of occluded GO sheets on the nanometer scale, we further employed scanning-TEM (STEM) coupled with a high-angle annular dark-field imaging (HAADF) detector, thus enabling imaging with chemical contrast. HAADF-STEM images obtained for the hybrid GO/calcite crystals elegantly revealed a layered distribution of the GO sheets, imaged here as darker stripes owing to their lower atomic number (Z) than that of the calcite matrix (Figure 2B, arrows). Imaging of the control calcite by the same technique revealed no such contrast (Figure 2C). The striped distribution of GO sheets observed on the HAADF-STEM images of the GO/calcite sample correlates well with the layered structure of the hybrid crystal edges observed on HRSEM (Figure 1D).

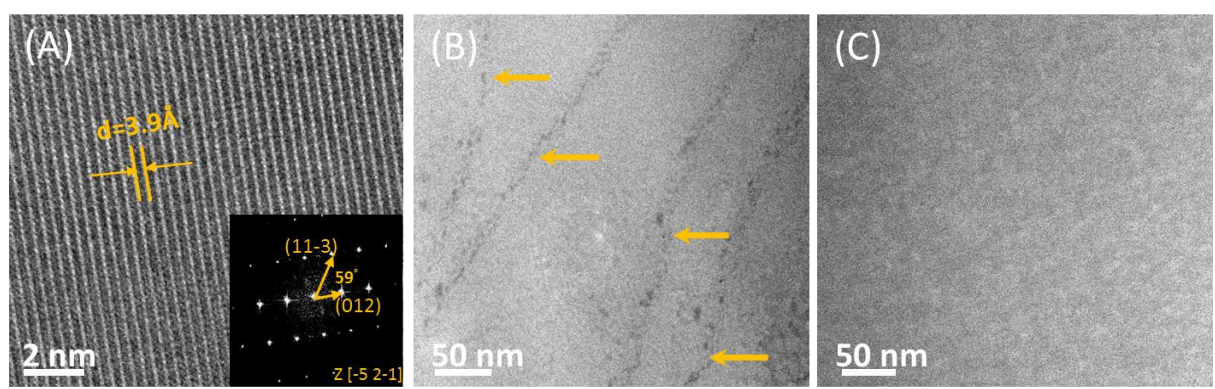

**Figure 2.** (A) HRTEM image of GO/calcite thin sample prepared by FIB and the coresponding fast Fourier transform (FFT) over the entire region. (B) HAADF-STEM image of calcite with embedded GO sheets appearing as dark stripes (see indicating arrows), and of pure calcite showing no Z-contrast (C).

Incorporation of various organic molecules into an inorganic host is widely accepted to lead to lattice distortions and unique microstructures.[13] Therefore, to determine plausible effects of the inclusion of GO on the calcite crystal lattice we performed high-resolution synchrotron powder X-ray diffraction (HRPXRD) at a dedicated synchrotron beam line (ID22 of the European Synchrotron Research Facility (ESRF), Grenoble, France) at a wavelength of 0.39970(8)Å. Both the control and the composite crystals exhibited a pure calcite phase (Figure 3A). Diffraction patterns revealed considerable structural differences between the lattice parameters of GO/calcite and those of the pure $CaCO_3$. The observed shift in XRD peak positions towards smaller $2\theta$-values, shown here only for the {111} reflection (Figure 3B), strongly supports incorporation of GO into the calcite crystal lattice. Similar behavior associated with incorporation phenomena was previously demonstrated on different synthetic calcite systems occluding various types of organic matter.[12, 27, 45, 46]

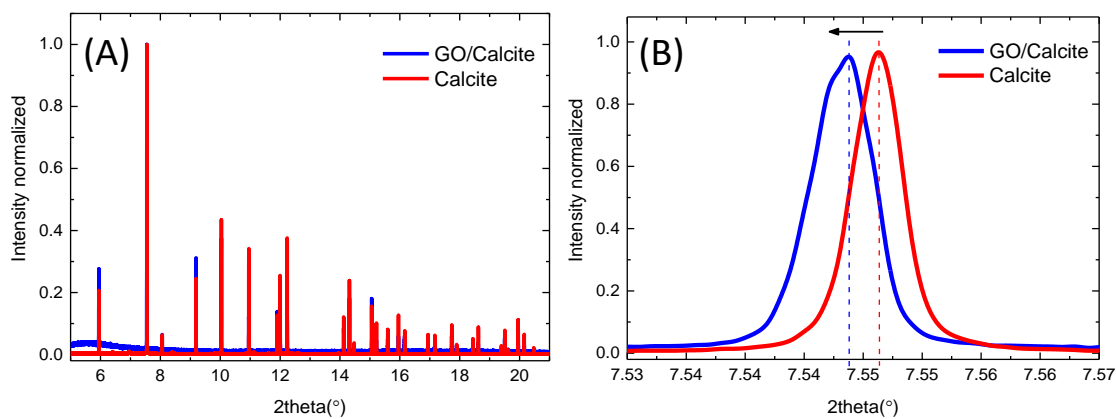

**Figure 3.** (A) HRPXRD patterns of pure calcite (red line) and hybrid GO/calcite (blue line) crystals. (B) Difference between the {104} reflection of the hybrid GO/calcite crystal relative to the control sample.

Rietveld structure refinement of the experimental data enabled us to precisely extract the lattice parameters and estimate the magnitude of the lattice distortions produced after entrapment of the GO sheets. The observed increase in the $c$-parameter of the hybrid calcite crystal resulted in positive lattice distortions of 4.23E-04 relative to the pure $CaCO_3$. By contrast, the slight decrease in $a,b$-parameters led to negative lattice distortions with a significantly lower magnitude of $-9.32$E-05 (Table 1).

**Table 1.** Lattice parameters, induced lattice distortions, unit cell volume and goodness of fit parameter.

| Sample | $a$, $b$-parameter, [Å] | Distortions $a,b$-axes | $c$-parameter, [Å] | Distortions $c$-axis | Unit cell volume, [Å³] | $\chi^2$ |
|---|---|---|---|---|---|---|
| Calcite | 4.990914(6) | - | 17.065653(8) | - | 368.14 | 2.259 |
| GO/calcite | 4.990449(4) | $-9.32$E-05 | 17.072865(5) | 4.23E-04 | 368.23 | 1.606 |

We next investigated the emission properties of GO upon its inclusion within the calcite crystals. Figure 4A shows the emission spectrum of a GO/calcite hybrid single crystal. The spectrum fits well to the GO emission reported in aqueous solvents.[38, 47] Notice that the emission spectrum was recorded locally on the crystal collecting the emission that arises from a section of ≈2 µm in diameter and that it did not change significantly when collected from different points on the crystal. Using time-correlated single photon counting (TCSPC), we also investigated the excited-state decay upon pulsed laser excitation at 405 nm. A multi-exponential decay was obtained, as shown in Figure 5B. The trace was fitted with a tri-exponential model that derived lifetimes of $\tau_1$=1.0 ns (48%), $\tau_2$=9.3 ns (13%), $\tau_3$=3.8 ns (39%), and hence an average lifetime of 3.2 ns. The 9.3 ns component of the decay is considerably longer than that previously observed for GO, which was in the range of picoseconds to a maximum of 2 ns.[38] Such an increase in the decay time might result from either the effect of the crystal's local polarity or the matrix rigidity that slows down the non-radiative deactivation rates of GO.[38-44]

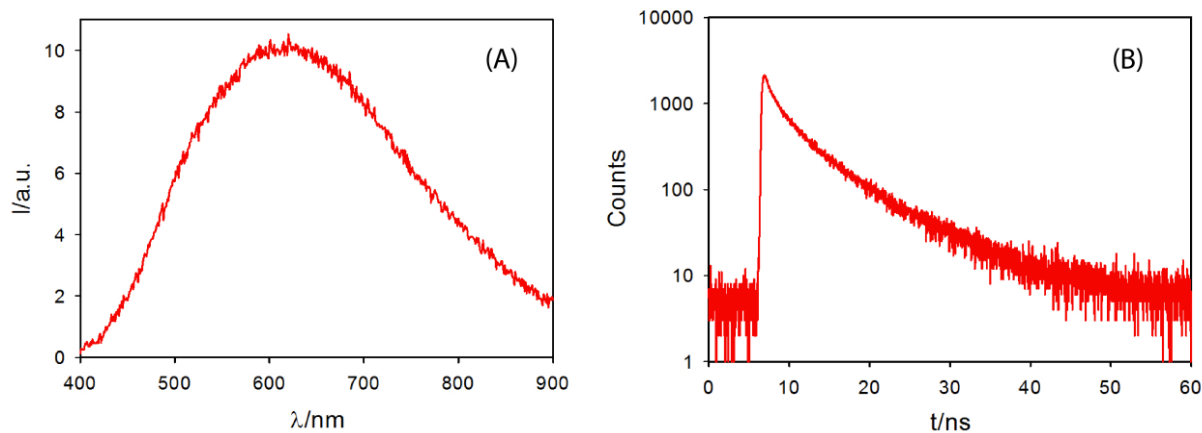

**Figure 4.** (A) Local fluorescence spectrum of GO-doped calcite upon excitation at 405 nm. (B) Excited-state decay of fluorescence in a GO-doped calcite crystal.

Transmission optical and fluorescence images of calcite crystals grown in the absence and in the presence of GO are compared in Figure 5. In the fluorescence detection mode pure calcite shows almost no signal (Figure 5D), whereas fluorescence is clearly detectable in the GO/calcite hybrid single crystal (Figure 5B). We performed these experiments utilizing an electron-multiplying charge coupled device (EMCCD) camera for image acquisition. A steady-state fluorescence spectroscope and a TCSPC apparatus were coupled with the optical microscope to record the local emission spectrum of GO in the crystals.

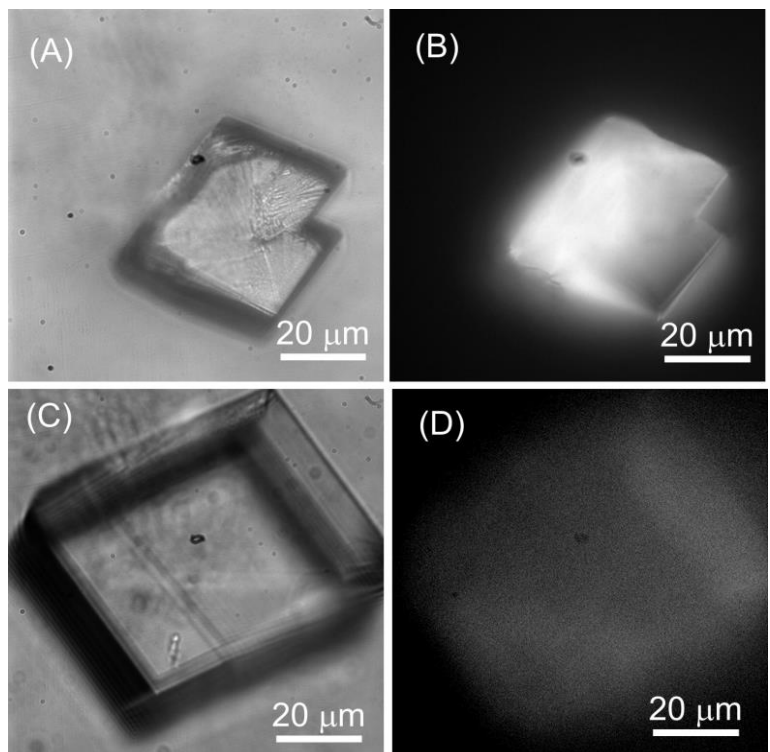

**Figure 5.** Transmission optical images of calcite crystals grown in the presence (A) and in the absence (C) of GO. On the right (B and D) are the corresponding transmission fluorescence images upon excitation at 405 nm.

These fluorescence imaging experiments again clearly indicate that GO is incorporated into the single crystal of calcite, and show that GO confers a new property on the calcite host. An optical micrograph (Figure S1A) and its corresponding crossed polarizer optical micrograph (Figure S1B) of several calcium carbonate single crystals precipitated in the presence of GO are observed. This experiment is yet another proof of the single crystallinity nature of the hybrid crystals.

Finally, we examined the mechanical properties of GO/calcite single crystals by nanoindentation testing[48] and analyzed the findings in comparison to those of control calcite crystals. To determine the elastic modulus ($E_{IT}$) and hardness ($H_{IT}$) as functions of the indentation depth, we analyzed the load/depth oscillations superimposed on the loading portion of the curve. Maximum indentation depths were measured at 200, 300, or 400 nm. From the data reported in Table 2, it appears that the two samples exhibited different values of $E_{IT}$ and $H_{IT}$. The presence of GO in the crystalline lattice produced an average decrease in $E_{IT}$ of about 12.5% and an increase in $H_{IT}$ of about 12.8%. This change was not affected by the indentation depth. Similar trends of decrease in elastic modulus and increase in hardness have been observed in an artificial biomineral.[10]

**Table 2.** $E_{IT}$ and $H_{IT}$ values calculated at indentation depths of 200, 300 or 400 nm, averaged over a set of five different crystals (five control and five hybrid crystals). Recorded are the percentages of variation ($\Delta\%$) in $E_{IT}$ and in $H_{IT}$ from calcite to GO/calcite at the different indentation depths. The uncertainty reported in the table is the standard deviation.

| Material | $E_{IT}$ (GPa) | | | $H_{IT}$ (GPa) | | |
|---|---|---|---|---|---|---|
| | 200 | 300 | 400 | 200 | 300 | 400 |
| GO/calcite | 59.3±4.5 | 55.7±3.4 | 53.3±3.7 | 4.40±0.84 | 4.12±0.85 | 4.16±0.89 |
| Calcite | 67.3±2.7 | 63.9±2.3 | 61.2±2.4 | 3.82±0.42 | 3.70±0.49 | 3.72±0.53 |
| $\Delta\%$ | −11.9 | −12.8 | −12.8 | 15.0 | 11.4 | 11.9 |

The reduction in elastic modulus can be explained by the fact that the GO layers are well oriented along the interface perpendicular to the direction of indentation (along the cleavage planes). It is therefore the out-of-plane Young's modulus, $E_{33}$, that has the main influence. Since the $E_{33}$ of graphite is smaller than that of calcite, and because the GO is incorporated in the direction perpendicular to the indentation, thus forming an alternating series of calcite and GO bundles, a softening of calcite is indeed expected. Accordingly, applying an inverse rule of mixture for the Young's modulus, as imposed by our in-series stacking ($1/E_{(Calcite/GO)} = 1/E_{(Calcite)} * (1-f) + 1/E_{(GO)} * f$), and assuming a minimum GO content of f = 0.027 in volume fraction (as deduced from the TEM image in Figure 2B), we deduce that $E_{33}$ for GO is 11.2, 9.9, and 9.4 GPa at 200, 300, and 400 nm of indentation, respectively.

On the other hand, the increase in hardness of the GO/calcite composite might suggest that both GO and the strain fields in the surrounding calcite lattice provide an obstacle to the motion of deformation twinning and the dislocation-mediated propagation of plastic deformation in the crystalline slip system. Moreover, applying the general shape/size-effect law for nanoindentation proposed in Pugno et al.,[49] we obtain

$$H(h) = H_0 * (1 + h_0/(h + h_1))^{(1/2)} \tag{1}$$

where H is the hardness, h the indentation depth, $H_0$ the macroscopic hardness, and $h_0$ and $h_1$ the characteristic lengths. For $H_0$ we find 3.8 GPa for the GO/calcite and 3.6 GPa for the calcite. We also derive $h_0 = 63.6$ nm and $h_1 \cong 0$ for GO/calcite and $h_0 = 28.8$ nm and $h_1 \cong 0$ ($R^2 \cong 1$) for pure calcite.

## 3. Conclusions

Our results showed that GO sheets can indeed become incorporated into calcite single crystalline hosts, allowing for the fabrication of graphene-based composite materials with enhanced properties. We demonstrated a layered distribution of GO sheets entrapped within the calcite host. These hybrid single crystals show new optical properties: in contrast to pure calcite, the hybrid crystals become fluorescent and are spectroscopically characterized by the presence of three lifetimes, one of which is considerably longer than that observed for bare GO in solution. Moreover, the mechanical properties of the calcite host are manipulated upon introduction of a carbon-based nanomaterial into its lattice. Compared to pure calcite, the composite GO/calcite crystals exhibit lower elastic modulus and higher hardness.

This study thus demonstrates that by incorporating 2D materials within a 3D crystal lattice we can obtain hybrid crystals possessing several new properties.

## 4. Experimental Section

*Materials*. Graphene oxide was purchased from Cheap Tubes Inc. (USA) and was used as supplied (lateral dimension 300−800 nm, thickness 0.7−1.2 nm, purity 99 wt%). Cao et al.[50] characterized this source of GO by X-ray photoelectron spectroscopy (XPS) and found that the carbon-to-oxygen ratio is approximately 4:1. The functional composition of the bulk GO was determined as follows: sp2 carbon (71.4%), hydroxyl groups (19.3%), carboxyl groups (9.3%). Dihydrate calcium chloride and granular anhydrous calcium chloride (for drying) were from Merck. Ammonium carbonate was supplied by Sigma-Aldrich. Solutions were prepared using water of Millipore grade (18.2 M$\Omega$).

*Synthesis.* The pH-controllable dispersion of GO is favored mainly by alkaline conditions. Whereas the presence of $Ca^{2+}$ facilitates GO aggregation, the dispersion of GO is assisted by ultrasonication. By optimization of these parameters, the optimal dispersion of GO was achieved in 10 mM $CaCl_2$ at pH 5.4 under ultrasonication for 60 min at 40% of the maximum amplitude of 120 $\mu$m (tip diameter 13 mm) and frequency of 20 kHz.

Calcite single crystals were precipitated at room temperature by controlled diffusion of $CO_2$ and $NH_3$ vapors into the 10 mM $CaCl_2$ solution containing previously dispersed GO in a concentration of 0.6 mg mL$^{-1}$. This simple method has the advantage of inducing slow nucleation and growth of calcite single crystals under the low supersaturation conditions that favor the additive incorporation.[51]

*Scanning Electron Microscopy.* A high-resolution scanning electron microscope (HR-SEM, ULTRA Plus, Zeiss, Oberkochen, Germany) was used to characterize the crystal shape and morphology. Samples were mounted on carbon tape and coated with carbon (Quorum Q150T ES, East Grinstead, UK). Images were obtained in high vacuum mode at 4 kV.

*Transmission and scanning transmission electron microscopy.* This was done with a Titan 80-300 FEG-S/TEM (FEI) microscope coupled with a HAADF detector. The microscope was operated at 300kV. The sample analyzed by STEM and TEM analysis was prepared utilizing an FEI Strata400S Dual-Beam focused ion beam (FIB).

*High-resolution Powder X-ray Diffraction.* HRPXRD analysis was carried out on the ID22 beamline of the ESRF. Powder samples were investigated at a wavelength of 0.39970(8)Å. All samples were measured in borosilicate 1 mm glass capillaries. Crystal lattice parameters were assessed by the Rietveld refinement method with GSAS software and the EXPGUI interface.[52, 53]

*Fluorescence microscopy.* Entrapping of GO within the calcite crystals causes changes in the spectroscopic properties of the hybrid crystals. The emission properties of GO upon its inclusion in calcite crystals were investigated by fluorescence microscopy using an inverted optical microscope IX-71 and a xenon lamp for excitation. An electron multiplying charge coupled device (EMCCD) camera (Princeton Instruments, PhotonMAX 512) was used for imaging acquisition. A steady state fluorescence spectroscope and a time-correlated single photon counting (TCSPC) apparatus were coupled to the microscope to record the local emission spectrum of GO in the crystals.

*Optical microscopy.* Leica transmission optical microscopy was used in order to obtain images of $CaCO_3$ crystals grown in the presence of GO. Samples were placed on a microscope slide beneath a standard glass coverslip and observed under bright-field conditions with crossed polarizers. Images were captured with a CCD digital camera and recorded using the LAS EZ software supplied by Leica Microsystems©.

*Nanoindentation analysis.* The mechanical properties of single crystals of pure calcite and of GO/calcite were measured with a nanoindentation tester model TTX-NHT (CSM Instruments), equipped with a Berkovich diamond tip and operating in continuous stiffness mode. Measurements were taken up to a maximum applied load of 30 mN. Using Oliver and Pharr dynamic analysis of the loading portion of the curve, we determined the instrumented (IT) values of the elastic Young's modulus ($E_{IT}$) and hardness ($H_{IT}$) as well as the stiffness S as functions of indentation depth.[38] Crystals were embedded at room temperature using fast curing, cold polymerizing resins based on methyl methacrylate (Technovit 5071 (Buehler)), and were then lightly polished using colloid alumina, average size 1 μm (PACE Technologies) to yield a clean, smooth, flat surface for indentation.

## Supporting information

Supporting information is available from the Wiley Online Library or from the authors.

## Acknowledgements

The research leading to these results received funding from the European Research Council under the European Union's Seventh Framework Program (FP/2007-2013)/ERC Grant Agreement [number 336077]. We are also indebted to the ESRF (ID22), and specifically to Dr. Andy Fitch, for use and support of the high-resolution powder beamline.

[+] M.D.G. and I.P. contributed equally to this work.

Received: ((will be filled in by the editorial staff))
Revised: ((will be filled in by the editorial staff))
Published online: ((will be filled in by the editorial staff))Copyright WILEY-VCH Verlag GmbH & Co. KGaA, 69469 Weinheim, Germany, 2013.

## References

[1]      H. A. Lowenstam, S. Weiner, *On biomineralization*, Oxford Univ. Press, New York **1989**.
[2]      L. Addadi, S. Weiner, *Angew. Chem. Int.Edit.* **1992**, 31, 153.
[3]      S. Mann, *Biomineralization: Principles and Concepts in Bioinorganic Materials Chemistry* Oxford University Press,  **2001**.
[4]      P. Fratzl, H. S. Gupta, F. D. Fischer, O. Kolednik, *Adv. Mater.* **2007**, 19, 2657.
[5]      K. Tai, M. Dao, S. Suresh, A. Palazoglu, C. Ortiz, *Nat. Mater.* **2007**, 6, 454.
[6]      F. Nudelman, N. A. J. M. Sommerdijk, *Angew. Chem. Int.Edit.* **2012**, 51, 6582.
[7]      S. Kim, C. B. Park, *Adv. Funct. Mater.* **2013**, 23, 10.
[8]      L. A. Estroff, H. Y. Li, M. Kunitake, H. L. L. Xin, A. Vodnick, S. P. Baker, D. A. Muller, *Abstr. Pap. Am. Chem. S* **2009**, 237.
[9]      M. E. Kunitake, L. M. Mangano, J. M. Peloquin, S. P. Baker, L. A. Estroff, *Acta Biomater.* **2013**, 9, 5353.
[10]   Y. Y. Kim, K. Ganesan, P. Yang, A. N. Kulak, S. Borukhin, S. Pechook, L. Ribeiro, R. Kroger, S. J. Eichhorn, S. P. Armes, B. Pokroy, F. C. Meldrum, *Nat. Mater.* **2011**, 10, 890.
[11]   A. Berman, L. Addadi, S. Weiner, *Nature* **1988**, 331, 546.
[12]   B. Pokroy, A. N. Fitch, P. L. Lee, J. P. Quintana, E. N. Caspi, E. Zolotoyabko, *J. Struct. Biol.* **2006**, 153, 145.
[13]   E. Weber, B. Pokroy, *Crystengcomm.* **2015**, 17, 5873.
[14]   S. Frølich, H. O. Sørensen, S. S. Hakim, F. Marin, S. L. S. Stipp, H. Birkedal, *Cryst. Growth Des.* **2015**, 15, 2761.

[15]    B. Pokroy, A. N. Fitch, E. Zolotoyabko, *Adv. Mater.* **2006**, 18, 2363.

[16]    B. Pokroy, A. N. Fitch, E. Zolotoyabko, *Cryst. Growth Des.* **2007**, 7, 1580.

[17]    S. Borukhin, L. Bloch, T. Radlauer, A. H. Hill, A. N. Fitch, B. Pokroy, *Adv. Funct. Mater.* **2012**, 22, 4216.

[18]    A. S. Schenk, I. Zlotnikov, B. Pokroy, N. Gierlinger, A. Masic, P. Zaslansky, A. N. Fitch, O. Paris, T. H. Metzger, H. Cölfen, P. Fratzl, B. Aichmayer, *Adv. Funct. Mater.* **2012**, 22, 4668.

[19]    A. N. Kulak, M. Semsarilar, Y. Y. Kim, J. Ihli, L. A. Fielding, O. Cespedes, S. P. Armes, F. C. Meldrum, *Chem. Sci.* **2014**, 5, 738.

[20]    Y. J. Liu, W. T. Yuan, Y. Shi, X. Q. Chen, Y. Wang, H. Z. Chen, H. Y. Li, *Angew. Chem. Int. Edit.* **2014**, 53, 4127.

[21]    C. Yao, A. Xie, Y. Shen, J. Zhu, H. Li, *Mater. Sci. Eng. C* **2015**, 51, 274.

[22]    M. Calvaresi, G. Falini, S. Bonacchi, D. Genovese, S. Fermani, M. Montalti, L. Prodi, F. Zerbetto, *Chem. Commun. (Camb)* **2011**, 47, 10662.

[23]    M. Calvaresi, G. Falini, L. Pasquini, M. Reggi, S. Fermani, G. C. Gazzadi, S. Frabboni, F. Zerbetto, *Nanoscale* **2013**, 5, 6944.

[24]    Y.-Y. Kim, M. Semsarilar, J. D. Carloni, K. R. Cho, A. N. Kulak, I. Polishchuk, C. T. Hendley, P. J. M. Smeets, L. A. Fielding, B. Pokroy, C. C. Tang, L. A. Estroff, S. P. Baker, S. P. Armes, F. C. Meldrum, *Adv. Funct. Mater.* **2016**, n/a.

[25]    X. Huang, X. Qi, F. Boey, H. Zhang, *Chem. Soc. Rev.* **2012**, 41, 666.

[26]    A. C. Ferrari, F. Bonaccorso, V. Fal'ko, K. S. Novoselov, S. Roche, P. Boggild, S. Borini, F. H. L. Koppens, V. Palermo, N. Pugno, J. A. Garrido, R. Sordan, A. Bianco, L. Ballerini, M. Prato, E. Lidorikis, J. Kivioja, C. Marinelli, T. Ryhanen, A. Morpurgo, J. N. Coleman, V. Nicolosi, L. Colombo, A. Fert, M. Garcia-Hernandez, A. Bachtold, G. F. Schneider, F. Guinea, C. Dekker, M. Barbone, Z. P. Sun, C. Galiotis, A. N. Grigorenko, G. Konstantatos, A. Kis, M. Katsnelson, L. Vandersypen, A. Loiseau, V. Morandi, D. Neumaier, E. Treossi, V. Pellegrini, M. Polini, A. Tredicucci, G. M. Williams, B. Hong, J. H. Ahn, J. M. Kim, H. Zirath, B. J. van Wees, H. van der Zant, L. Occhipinti, A. Di Matteo, I. A. Kinloch, T. Seyller, E. Quesnel, X. L. Feng, K. Teo, N. Rupesinghe, P. Hakonen, S. R. T. Neil, Q. Tannock, T. Lofwander, J. Kinaret, *Nanoscale* **2015**, 7, 4598.

[27]    E. Weber, L. Bloch, C. Guth, A. N. Fitch, I. M. Weiss, B. Pokroy, *Chem. Mater.* **2014**, 26, 4925.

[28]    X. Huang, X. Z. Zhou, S. X. Wu, Y. Y. Wei, X. Y. Qi, J. Zhang, F. Boey, H. Zhang, *Small* **2010**, 6, 513.

[29]    Y. T. Kim, J. H. Han, B. H. Hong, Y. U. Kwon, *Adv. Mater.* **2010**, 22, 515.

[30]    S. Kim, S. H. Ku, S. Y. Lim, J. H. Kim, C. B. Park, *Adv. Mater.* **2011**, 23, 2009.

[31]    P. A. A. P. Marques, G. Goncalves, M. K. Singh, J. Gracio, *J. Nanosci. Nanotechno.* **2012**, 12, 6686.

[32]    G. M. Neelgund, A. Oki, Z. P. Luo, *Mater. Res. Bull.* **2013**, 48, 175.

[33]    S. Bai, X. P. Shen, *Rsc. Adv.* **2012**, 2, 64.

[34]    X. Wang, H. Bai, Y. Jia, L. Zhi, L. Qu, Y. Xu, C. Li, G. Shi, *Rsc. Adv.* **2012**, 2, 2154.

[35]    A. E. Nielsen, *Pure Appl. Chem.* **1981**, 53, 2025.

[36]    P. Johari, V. B. Shenoy, *ACS Nano* **2011**, 5, 7640.

[37]    J. W. Suk, R. D. Piner, J. H. An, R. S. Ruoff, *ACS Nano* **2010**, 4, 6557.

[38]    J. Z. Shang, L. Ma, J. W. Li, W. Ai, T. Yu, G. G. Gurzadyan, *Sci. Rep-Uk.* **2012**, 2.

[39]    C. T. Chien, S. S. Li, W. J. Lai, Y. C. Yeh, H. A. Chen, I. S. Chen, L. C. Chen, K. H. Chen, T. Nemoto, S. Isoda, M. W. Chen, T. Fujita, G. Eda, H. Yamaguchi, M. Chhowalla, C. W. Chen, *Angew. Chem. Int.Edit.* **2012**, 51, 6662.

[40]    S. K. Cushing, M. Li, F. Huang, N. Wu, *ACS Nano* **2014**, 8, 1002.

[41]    G. Eda, Y. Y. Lin, C. Mattevi, H. Yamaguchi, H. A. Chen, I. S. Chen, C. W. Chen, M. Chhowalla, *Adv. Mater.* **2010**, 22, 505.

[42]    A. L. Exarhos, M. E. Turk, J. M. Kikkawa, *Nano Lett.* **201**, 13, 344.

[43]    D. Kozawa, Y. Miyauchi, S. Mouri, K. Matsuda, *J. Phys. Chem. Lett.* **2013**, 4, 2035.

[44]    S. Vempati, T. Uyar, *Phys. Chem. Chem. Phys.* **2014**, 16, 21183.

[45]    B. Pokroy, J. P. Quintana, E. N. Caspi, A. Berner, E. Zolotoyabko, *Nat. Mater.* **2004**4, 3, 900.

[46]    G. Magnabosco, M. Di Giosia, I. Polishchuk, E. Weber, S. Fermani, A. Bottoni, F. Zerbetto, P. G. Pelicci, B. Pokroy, S. Rapino, G. Falini, M. Calvaresi, *Adv. Healthc. Mater.* **2015**, 4, 1510.

[47]    K. P. Loh, Q. Bao, G. Eda, M. Chhowalla, *Nat. Chem.* **2010**, 2, 1015.

[48]    W. C. Oliver, G. M. Pharr, *J. Mater. Res.* **1992**, 7, 1564.

[49]    N. M. Pugno, *Acta Mater.* **2007**, 55, 1947.

[50]    C. Cao, M. Daly, C. V. Singh, Y. Sun, T. Filleter, *Carbon* **2015**, 81, 497.

[51]    J. Ihli, P. Bots, A. Kulak, L. G. Benning, F. C. Meldrum, *Adv. Funct. Mater.* **2013**, 23, 1965.

[52]    B. Toby, *J. App. Crystallogr.* **2001**, 34, 210.

[53]    A. C. Larson, R. B. Von Dreele, *Los Alamos National Laboratory Report LAUR* **2000**, 86.